# Spontaneous and stimulated electron-photon interactions in nanoscale plasmonic near fields

Matthias Liebtrau[1], Murat Sivis[2,3], Armin Feist[2], Hugo Lourenço-Martins[2], Nicolas Pazos-Pérez[4], Ramon A. Alvarez-Puebla[4,5], F. Javier García de Abajo[5,6], Albert Polman[1*], and Claus Ropers[2,3]

[1]Center for Nanophotonics, AMOLF, 1098 XG Amsterdam, The Netherlands
[2]4th Physical Institute, Solids and Nanostructures, University of Göttingen, 37077 Göttingen, Germany
[3]Max Plank Institute for Biophysical Chemistry, 37077 Göttingen, Germany
[4]Department of Physical Chemistry and EMaS, Universitat Rovira I Virgili, 43007 Tarragona, Spain
[5]ICREA, 08010 Barcelona, Spain
[6]ICFO, 08860 Castelldefels, Barcelona, Spain

**Abstract**

We demonstrate spatially-resolved measurements of spontaneous and stimulated electron-photon interactions in nanoscale optical near fields using electron energy-loss spectroscopy (EELS), cathodoluminescence spectroscopy (CL), and photon-induced near-field electron microscopy (PINEM). Specifically, we study resonant surface plasmon modes that are tightly confined to the tip apexes of an Au nanostar, enabling a direct correlation of EELS, CL, and PINEM on the same physical structure at the nanometer length scale. Complemented by numerical electromagnetic boundary-element method calculations, we discuss the spontaneous and stimulated electron-photon interaction strength and spatial dependence of our EELS, CL and PINEM distributions. We demonstrate that in the limit of an isolated tip mode, spatial variations in the electron-near field coupling are fully determined by the modal electric field profile, irrespective of the spontaneous (in EELS and CL) or stimulated nature (in PINEM) of the process. Yet we show that coupling to the tip modes crucially depends on the incident electron energy with a maximum at a few keV, depending on the proximity of the interaction to the tip apex. Our results provide elementary insights into spontaneous and stimulated electron-light-matter interactions at the nanoscale that have key implications for research on (quantum) coherent optical phenomena in electron microscopy.

Nanoscale optical components enable light manipulation at deep-subwavelength length scales with a broad variety of applications in quantum information systems, optical signal processing, photovoltaics, molecular sensing, chemical-catalysis, and more.[1] The small feature sizes rendering the unique optical properties of these structures demand novel optical characterization techniques that overcome the diffraction-limited resolution of traditional light microscopy. In the recent past, scanning (transmission) electron microscopy (S(T)EM) has been established as a powerful platform to probe optical material properties with high spatial, temporal and energy resolution.[2–4]

Swift electrons with kinetic energies in the 1-300 keV range drive optical materials excitations in a unique way.[2] Similar to an optical pulse, the evanescent electric field of a free electron coherently couples to oscillations of the polarizable charges in a material. As the electron passes by, these charges experience a single electromagnetic field cycle with a duration of a few hundred attoseconds, corresponding to an excitation energy spectrum with significant weights between zero and several tens of eV.[5] The electron thus serves as an ultra-broadband excitation source that can couple to radiative and non-radiative modes in a material over the full ultraviolet-visible-infrared (UV-VIS-IR) spectral range. Electron energy-loss spectroscopy (EELS) reveals the energy transfer upon excitation of an optical resonance by measuring the energy loss experienced by the electron.[6,7] Additionally, cathodoluminescence (CL) spectroscopy unveils the material's emission characteristics by detecting radiative charge relaxation in the far field.[8]

The spatial and spectral characteristics of the material resonances are directly reflected in the measured EELS and CL signals, rendering the electron beam a highly versatile near-field probe. In fact, the spontaneous electron energy-loss and photon-emission probabilities are closely linked to the full and radiative electromagnetic local density of states (EMLDOS), respectively.[9,10] Thus, EELS probes both the dark and bright modes in a material, while CL is sensitive to the bright modes only.[11] Due to the coherent nature of the electron excitation mechanism, the two techniques are ideally suited for correlated structural and optical characterization of plasmonic and dielectric nanoparticles,[12–17] optical waveguides,[18–20] photonic crystal cavities,[21,22] and more.[3,4,7]

Recently, EELS and CL have been complemented with photon-induced near-field electron microscopy (PINEM).[23] In this technique, swift electrons are used to probe the near field of a nanostructure excited by an intense laser field. While passing through this near field, the electrons undergo one or multiple energy-gain and -loss transitions by stimulated absorption and emission of photons at the laser frequency $\omega_L$.[24–26] As a consequence, the initial electron energy spectrum is expanded with sidebands, evenly spaced by the photon energy $\hbar\omega_L$. The population of these sidebands varies with the intensity of the laser-induced optical field and the statistics of the excitation,[27] enabling near-field measurements with extreme spatial, temporal, and energy resolution.[28–33] Key to the PINEM mechanism is the fact that the evanescent near field provides spatial Fourier components with sufficiently large momenta to bridge the momentum mismatch between electrons (i.e., the electron electric field) and the driving field in free space. For large incident light intensities, the PINEM interaction can be a highly efficient process in which nearly every electron undergoes stimulated energy-gain or -loss transitions,[29] even leading to hundreds of net photon exchanges.[34,35] In contrast, in EELS and CL the probability for the electron to spontaneously drive an optical excitation is small, typically in the order of $10^{-5}$-$10^{-3}$ per eV energy bandwidth in the visible spectral range.[2]

The full exploitation of the rich new physics that the PINEM effect offers is just starting.[3,4,27–39] Recent research has focused on studying the quantum nature of the electron wave packet during the interaction with an optical near field and the subsequent modulation of the electron wave packet, enabling exciting phenomena such as the generation of coherent attosecond electron pulse trains[31,36,37] or electron vortex beams.[38] The relation between the stimulated and spontaneous interaction mechanisms governing PINEM, EELS, and CL has been addressed theoretically for small dipolar particles.[40,41] Yet an experimental comparison of the three techniques on a single complex structure has not been reported to date.

In this article, we present spatially-resolved EELS, CL, and PINEM measurements on a single Au nanostar with sharp conical tips.[42] As shown in previous works,[43–52] these tips sustain distinct plasmonic resonances that give rise to highly confined optical near fields at the tip apexes, providing an ideal geometry to correlate EELS, CL, and PINEM measurements at the nanometer length scale. Supported by theoretical considerations and numerical electromagnetic boundary-element method (BEM) calculations, we test the hypothesis that in the limit of an isolated tip mode spatial variations in the electron-near field coupling are fully determined by the modal electric field profile, irrespective of whether the excitations are driven by the electron itself (in EELS and CL) or a laser pulse (in PINEM). We discuss the link between the electron-near field coupling strength and the electron energy, and show its dependence on the spatial Fourier composition of the optical field. Our data demonstrate the signal quality and spatial resolution that is practically achieved by state-of-the art EELS, CL, and PINEM instruments, given experimental limitations and uncertainties. We provide detailed insights into the correlation between spontaneous and stimulated electron-near field interactions, yielding a starting point for further combined EELS, CL, and PINEM experiments to explore new quantum phenomena in electron-light-matter interactions.

**Results and Discussion**

**Electron-Near Field Interactions.** In this work, spatially-resolved EELS, CL, and PINEM measurements are carried out on a single chemically-synthesized Au nanostar composed of an approximately 50-nm-diameter spherical core and sharp conical protrusions with tip-radii of curvature < 3 nm.[42] The structure is deposited on a thin electron-transparent silicon nitride support membrane. Assuming negligible spatial overlap and hence vanishing coupling between the tip resonances,[51] we can describe the plasmon-mediated electron-near field interaction by considering the optical response of a single Au tip attached to a spherical core. We assume a swift electron of initial energy $E_0$ propagating along the **z** direction and passing near the tip oriented along the **x** direction. In this configuration, the time-varying evanescent electric field of the electron couples most efficiently to the dominant in-plane dipole moment $\mathbf{p}_x$ along the symmetry axis of the tip. The $z$ component of the electric field $E_z$ associated with that $\mathbf{p}_x$ dipole acts back on the electron, resulting in an energy loss $\Delta E$ with a spectral probability distribution peaked around the tip resonance energy. Subsequently, the energy transferred to the particle is either dissipated as heat or radiated into the far field, giving rise to CL.

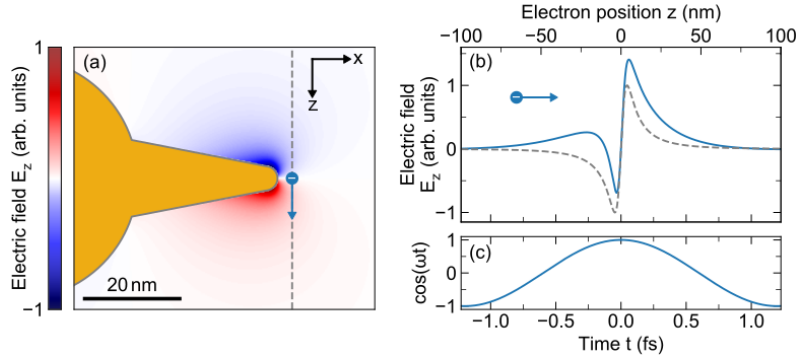

**Fig. 1** | BEM calculation for swift electron excitation of a conical Au nanotip attached to a spherical Au core (tip length 30 nm, cone aperture angle 22°, tip apex radius-of-curvature 2.5 nm, core diameter 50 nm). At an incident energy of 20 keV the electron passes 3 nm away from the tip along the dashed line perpendicular to the symmetry axis of the tip ($z$ direction). (a) Frozen-time snapshot at minimum electron-apex distance ($t$=0) of the $z$ component of the optical electric field $E_z^e$ induced upon excitation of the tip resonance at $\hbar\omega_0$ =1.73 eV. (b) Dashed-grey curve: $E_z^e$ profile along the electron trajectory in (a), with $z = 0$ corresponding to the tip symmetry axis. Solid-blue curve: time evolution of $E_z^e$ as experienced by the electron while propagating through the electron-self-induced plasmon field. (c) Time evolution of the optical phase of $E_z^e$, oscillating at the plasmon resonance frequency $\omega_0$.

To illustrate the spontaneous electron energy-loss mechanism, we numerically calculate the electric field induced by a 20 keV electron passing near the tip apex of the model geometry depicted in Fig. 1(a). Full-wave retarded boundary element method (BEM)[53] calculations are performed using the MNPBEM17 toolbox[54,55] with optical constants for Au taken from Ref. 56. The color map shows the real part of the induced $E_z$ distribution in the $x$-$z$ symmetry plane of the structure. The electron is incident from the top, travelling along the dashed-grey line as indicated by the blue arrow. The map represents a time snapshot of the spectral electric field component induced at the tip resonance energy of 1.73 eV for $t = 0$ (i.e., the moment of maximum approach between the electron and the tip). The field distribution is strongly localized near the tip apex, vanishing along the symmetry axis of the tip while showing opposite signs above and below (upward field orientation in red; downward field in blue).

In the no-recoil approximation ($\Delta E \ll E_0$), the spectral electron energy-loss probability can be calculated from the $z$ component of the electron-self-induced electric field $E_z^e$ according to[2]

$$\Gamma_{\text{loss}}(\mathbf{R}, \omega) = \frac{e}{\pi \hbar \omega} \, \text{Re} \left\{ \int_{-\infty}^{\infty} E_z^e(\mathbf{R}, z) e^{-i\frac{\omega}{v}z} dz \right\}, \tag{1}$$

where $\mathbf{R} = (x, y)$ and $z$ denote the lateral and along-the-beam electron positions, respectively, $e$ is the electron charge, and $v$ is the electron velocity. The exponential term describes temporal oscillations in the phase of $E_z^e$ as the electron passes by the tip. Here, $z = vt$ is linked to time $t$ via the electron velocity $v$. The solid-blue curve in Fig. 1(b) shows the full excursion of the real part of $E_z^e$ as experienced by the electron during its passage through the plasmon field. For reference, the dashed-grey curve shows the field profile for $t = 0$, and the time evolution of its optical phase is plotted in Fig. 1(c). As the electron passes by, the field changes sign multiple times, with the number of oscillations increasing with decreasing electron velocity $v$. Classically, this implies that the electron experiences subsequent acceleration and deceleration along its trajectory, leading to alternating positive and negative contributions to the energy-loss probability. In fact, according to the integral expression in Eq. (1), net energy exchange only occurs for the spatial Fourier field component that corresponds to a wave with phase velocity equal to $v$. The loss probability is then determined by the real part of the Fourier amplitude of $E_z^e$ at the electron velocity-dependent spatial frequency $q = \omega/v$.

Eq. (1) describes the backaction of the induced electric field on the electron, however, no information is provided on the excitation of the field. To this end, it is constructive to consider the decomposition of $E_z^e$ into the electric field eigenmodes $E_z^i$ of the structure.[9–11] As discussed in Refs. 10 and 11, for the highly localized plasmonic tip modes, the spectral electron energy-loss (EELS) and photon-emission (CL) probability densities related to the excitation of mode $i$ can be written as

$$\Gamma_{\text{loss},i}(\mathbf{R}, \omega) = \frac{e^2}{\pi \hbar \omega^2} \text{Im}\{f_i(\omega)\} |E_z^i(\mathbf{R}, q)|^2, \tag{2}$$

$$\Gamma_{\text{rad},i}(\mathbf{R}, \omega) = \frac{e^2 \omega}{4\pi^2 \hbar c^3} A_i |f_i(\omega)|^2 |E_z^i(\mathbf{R}, q)|^2, \tag{3}$$

where $f_i(\omega)$ and $A_i$, respectively, represent the spectral line shape and albedo of the mode, and $E_z^i(\mathbf{R}, q)$ denotes the complex Fourier amplitude of $E_z^i(\mathbf{R}, z)$ at $q = \omega/v$. Importantly, these expressions show that spatial variations in the electron energy-loss and photon-emission probabilities only depend on the electric field profile of the mode, independent of the electric field exerted by the electron itself. Their dependence with field intensity reflects that the electron-near field coupling persists both upon excitation of the mode and its backaction on the electron. This is consistent with the fact that the probability must be associated with the square of a matrix element coupling the electron and field modes, which in turn involves a spatial integral over $eE_z^i(\mathbf{R}, q)$. We note that for modes with spatial and spectral overlap, far-field interferences can contribute to the total photon emission probability $\Gamma_{\text{rad}}$.[10]

Next, we discuss the PINEM effect for the same geometry. If the laser polarization has a component oriented along the symmetry axis of the tip, it will drive the same $\mathbf{p}_x$ dipolar resonance as is excited by the electron in EELS and CL. However, for laser peak intensities in the order of hundreds of MW/cm² as used in this work the induced dipole strength is much larger than that generated by an individual electron. The interaction probability is thus strongly enhanced at the laser frequency $\omega_L$. Furthermore, the external driving permits electrons to gain energy, facilitating both energy-gain and -loss transitions by stimulated absorption or emission of photons at an energy exchange $\hbar\omega_L$. As demonstrated in Ref. 29, the energy spectrum of the transmitted electrons then evolves into a ladder of coherent energy-gain and -loss states, with the population of the ladder states governed by Rabi oscillations in the electron-light energy exchange process.

The probability for an electron to undergo a net amount of $n$ stimulated energy-gain or -loss transitions is given by Bessel functions of the first kind, $n^{th}$ order[25,29]

$$P_n(\mathbf{R}, \omega) = J_n^2(2|\beta(\mathbf{R}, \omega)|) \, \delta(\omega - \omega_L), \tag{4}$$

where

$$\beta = \frac{e}{\hbar\omega_L} E_z^L(\mathbf{R}, q) = \frac{e}{\hbar\omega_L} \int dz \, E_z^L(\mathbf{R}, z) \, e^{-iqz} \tag{5}$$

is the coupling coefficient of the electron to the time-varying laser-induced electric field $2\text{Re}\{E_z^L(\mathbf{R}, z)e^{-i\omega_L t}\}$ (the coupling coefficient $\beta$ is denoted $g$ elsewhere,[29,35] accompanied by a leading factor of $1/2$ and a different normalization of the time-varying field as $\text{Re}\{E_z^L(\mathbf{R}, z)e^{-i\omega_L t}\}$, without the leading factor of 2). As above for EELS and CL, coupling is only possible for a spatial Fourier field component at $q = \omega_L/v$. Decomposing $E_z^L$ into the electric field eigenmodes of the material, we can approximate the coupling strength of the electron to mode $i$ by an expression of the form

$$|\beta_i(\mathbf{R}, \omega)| = \frac{e}{\hbar\omega} \sqrt{\eta_i I} \sqrt{A_i} |f_i(\omega)| |E_z^i(\mathbf{R}, q)|, \tag{6}$$

where $\eta_i$ reflects the coupling efficiency of the driving field to the modal dipole moment $\mathbf{p}_i$ for a given angle of incidence and polarization, and $I$ is the field intensity. As can be seen, the squared coupling strength $|\beta_i|^2$ has the same spatial dependence as the spontaneous electron energy-loss and photon emission probability densities $\Gamma_{\text{rad},i}$ and $\Gamma_{\text{loss},i}$. This implies that for a given mode and at a fixed electron beam energy, EELS, CL, and PINEM yield equivalent spatial distributions. Furthermore, $|\beta_i|^2$ has the same dependence on $f_i(\omega)$ and $A_i$ as $\Gamma_{\text{rad},i}$, showing that both PINEM and CL are linked to the radiative modes and thus the far field emission characteristics of a material. Incidentally, it has been rigorously shown[27] that the coupling coefficient associated with a mode $i$ depends on its population $n_i$ as $|\beta_i(\mathbf{R}, \omega)| \propto \sqrt{n_i}$, which further corroborates the dependence shown in Eq. (6) on efficiency and intensity because $n_i \propto \eta_i I$.

**EELS and CL Experiments.** Spatially-resolved EELS and CL measurements are performed in STEM and SEM instruments operated at electron beam energies of 200 keV and 20 keV, respectively. The electron beam is raster-scanned over a two-dimensional grid of pixels with dimensions of (2×2) nm². Figure 2(a) shows EELS spectra acquired at four different tip apexes and the center of the nanostar core. At the tips, plasmonic resonances give rise to pronounced maxima at 1.48 eV, 1.80 eV, 1.84 eV, and 1.94 eV, with a full-width-at-half-maximum (FWHM) in the order of 400 meV. The core shows a broad, comparably flat spectrum, with another low-energy maximum near 1.0 eV. Figure 2(c) shows CL spectra acquired at approximately the same five positions. Again, the tip spectra indicate distinct plasmonic resonances, with maxima peaking at energies of 1.76 eV, 1.80 eV, 1.97 eV, and 2.04 eV, and an FWHM around 200 meV. In the core spectrum, we observe peaks of similar width, yet smaller amplitude, at 1.8 eV, 2.0 eV, and 2.3 eV.

As is evident from Fig. 2(a,b), there is significant spectral overlap between the features associated with the different tips and nanostar core. The high spectral resolution in CL permits to observe shoulders on the low- or high-energy side of the peaks, indicating that an electron exciting a single tip also drives the resonance of another tip. For example, the blue spectrum peaking at 1.73 eV shows a shoulder around 2.0 eV, in agreement with the positions of the peaks in the orange and purple spectra. Another shoulder shows up around 2.4 eV, matching well with the high-energy maximum in the core spectrum.

In good correspondence with the results of earlier experiments on Au nanostars,[48,50,51] we attribute the high-energy peak near 2.4 eV to the plasmon resonance of the nanostar core. In fact, this value agrees well with the calculated dipolar Mie resonance for a 50-nm-diameter Au sphere in vacuum. The side peaks in the core spectrum at 1.8 eV and 2.0 eV indicate coupling between the core and the tip modes as suggested in Ref. 45. We note that in EELS, we did not retrieve a sufficiently distinct plasmon signal near 2.4 eV to unambiguously identify the core resonance. We attribute this to multiple inelastic scattering upon electron transmission through the nanostar core and the thicker sections of the nanostar tips.

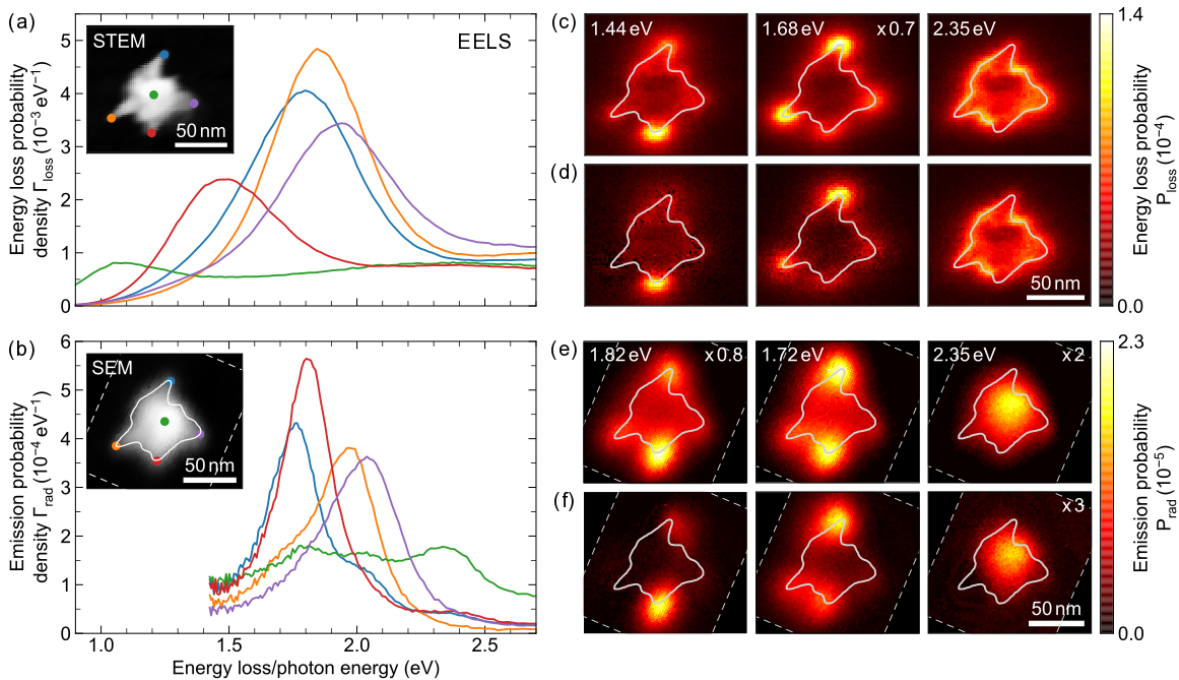

**Fig. 2 |** (a) 200 keV STEM-EELS and (b) 20 keV SEM-CL spectra taken at four different tips and at the core of the same chemically-synthesized Au nanostar. Electron beam positions are indicated by the color-matched dots in the insets, showing a STEM bright-field image and an SEM image of the nanostar, respectively. All spectra represent an average over 5×5 neighboring pixels. Energy-filtered (c) EELS and (e) CL probability distributions obtained for a bandwidth of ±25 meV around the plasmon resonance energies indicated on the top. Fitted (d) EELS and (f) CL probability distributions, revealing the extracted contribution of the tip plasmon resonances to the raw energy-filtered maps in (c) and (e). The boundaries of the original acquired data sets are indicated by the thin dashed lines in (b,e,f). The solid lines superimposed on the EELS and CL maps and the SEM image illustrate the approximate contour of the nanostar as inferred from the contrast in the STEM bright-field image in (a).

To gain more quantitative information and isolate the local contribution of individual plasmon resonances, we fit the EELS and CL data with a model assuming the nanostar response to be dominated by the resonances of the four tips marked by the colored dots in Fig. 2(a,b). In CL, we also take into account the core resonance. As mentioned above, In this approach, we neglect interference effects in the far field, assuming that there is vanishing spatial overlap between the modal fields.[10] In the past, a similar procedure has been applied to nanostars,[48] nanotriangles,[11] and branched nanostructures.[57] For EELS, resonances are represented by a sum of Gaussians, reflecting the approximate shape of the ZLP. A background associated with (multiple) inelastic scattering is modelled by a Gaussian error function, rising from zero to a constant amplitude at energies > 2.4 eV. Furthermore, we add a Gaussian centered between 2.3 eV and 2.9 eV to account for weak plasmonic contributions from the core and/or other tips. Another Gaussian describes the low energy feature observed in the core spectrum, with its position bound to energies below 1.3 eV. For CL, we describe the plasmon resonances by a sum of pseudo-Voigt distributions, capturing both their natural Lorentzian line shape as well as inhomogeneous broadening (i.e., due to electron-beam-induced carbon contamination). A constant background accounts for the emission of transition radiation or weak incoherent luminescence upon direct electron impact onto the nanostar. Further details on the spectral fitting procedure are provided in the methods section.

An overview on the obtained resonance energies and linewidths is given in Tab. I. The linewidths retrieved from the CL analysis range between 0.16 eV and 0.27 eV, corresponding to an average quality factor $Q \approx 9$. Larger linewidths are found from the EELS data, ranging between 0.33 eV and 0.44 eV. We attribute this difference to the lower energy resolution in EELS, which is mainly limited by the intrinsic energy spread in the electron beam. Yet, we note that the data were deconvolved with the ZLP measured upon electron transmission through the silicon nitride support membrane, enabling a substantial improvement in energy resolution (see methods).[58]

**Tab. I |** Fitted plasmon resonance energies ($E_0$) and FWHM linewidths ($\gamma$) derived from EELS and CL data (see Fig. 2(a,b)).

|  | EELS (200 keV) | | CL (20 keV) | |
| --- | --- | --- | --- | --- |
|  | $E_0$ (eV) | $\gamma$ (eV) | $E_0$ (eV) | $\gamma$ (eV) |
| Tip I (red) | 1.44 | 0.40 | 1.82 | 0.16 |
| Tip II (blue) | 1.68 | 0.36 | 1.72 | 0.27 |
| Tip III (orange) | 1.85 | 0.33 | 1.95 | 0.22 |
| Tip IV (purple) | 1.98 | 0.44 | 2.06 | 0.24 |
| Core (green) | -- | -- | 2.35 | 0.25 |

Comparing the resonance energies for each tip, we find a consistent blue shift between the values retrieved from CL and those retrieved from EELS. For tips II-IV minor blue shifts range between 3-6%, while for tip I we find a considerable deviation of 27%. Earlier work on silver nanotriangles has demonstrated spectral shifts between CL and EELS maxima associated with dissipation in the metal itself and the support substrate.[11,59] In Ref. 11 two opposite regimes were identified with increasing CL/EELS blue and red shifts below and above resonance energies of 1.6 eV and 1.8 eV, respectively. However, we note that in our experimental procedure, the CL measurements were taken last, after EELS and PINEM experiments and that consecutive CL measurements have shown the tip resonances to shift to the blue with every iteration. Thus, a plausible explanation for the CL/EELS blue shift is electron beam-induced carbon contamination[60] which alters the dielectric environment of the tips in the presence of the silicon nitride support film. Furthermore, tip I might have slightly deformed under prolonged near-resonant laser-beam excitation during PINEM acquisition. In fact, the tip resonances are highly sensitive to the exact tip morphology, with decreasing sharpness and aspect ratio resulting in blue shifts of tens to hundreds of meV.[42]

Using the results of the modal decomposition, we can now plot the field distributions associated with each resonance. Figure 2(c,d) shows energy-filtered EELS and CL maps integrated over a bandwidth of ±25 meV around the resonances of tips I and II, as determined from EELS and CL, respectively, as well as the core resonance (see labels on the top right). In the latter case, the core region is bright in CL while in EELS it remains mostly dark, showing that multiple inelastic scattering limits EELS to the thin regions of the sample. As expected, for the tip resonances we observe distinct hotspots near the tip apexes. However, in some cases more than one tip lights up, which is a consequence of spectral overlap between the modes.

To resolve the spectral ambiguity in the raw energy-filtered maps, we plot the local resonance amplitudes retrieved from the fitting procedure described above in Fig. 2(d,f). For reference, the EELS map at 2.35 eV shows the fitted background contribution. Upon first inspection, there is good agreement in the localization of the tip modes found by EELS and CL. Notably, this observation also applies to tip I with the largest deviation in the measured resonance energies. Supported by the results of electromagnetic BEM calculations, we will systematically compare the measured EELS and CL distributions together with our PINEM data later on.

**PINEM Experiments.** Spatially-resolved PINEM measurements are performed in the same STEM instrument as used for EELS, operated in ultrafast photoemission configuration at an electron energy of 200 keV. Sub-picosecond electron pulses are temporally synchronized with 3.4-ps optical pump pulses of 1.55 eV central photon energy that are incident near-normal to the sample plane. As in EELS and CL, the electron beam is raster-scanned over a two-dimensional grid of pixels with dimensions of (2×2) nm². The inset in Fig. 3(a) shows a STEM high-angle-annular-dark-field (HAADF) image of the nanostar prior to PINEM acquisition, with the white arrow indicating the laser polarization. We note that simultaneous structural imaging is strongly limited due to the low electron beam current in photoemission operation of our STEM instrument. The main panel in Fig. 3(a) shows PINEM spectra taken near the apex of tip II with an approximate resonance energy of 1.7 eV (as determined by EELS and CL), at distances of about 5 nm (blue) and 20 nm (orange) from the tip, respectively. At 20 nm distance, we observe a pronounced ZLP, with the first-order emission and absorption peaks ($\pm\hbar\omega_L$) clearly visible on both the energy-gain and -loss sides. In contrast, at 5 nm distance, the ZLP is fully depleted and the first-, second-, and third-order sidebands ($\pm n\hbar\omega_L$ for $n = 1, 2$, and $3$) are observed. As discussed above, this trend indicates increasing electron-near field coupling with decreasing separation from the tip apex. Each sideband and the ZLP have a linewidth of about 0.9 eV as primarily determined by the energy spread of the electron pulses. We find a larger spread than in EELS due to excess energy in the photoemission process (source excitation photon energy 3.1 eV) and space charge-related broadening of the electron energy distribution close to the tip emitter.[61,62] Yet, we note that a fundamental limit to the spectral resolution in PINEM is only imposed by the bandwidth of the employed laser system.[32]

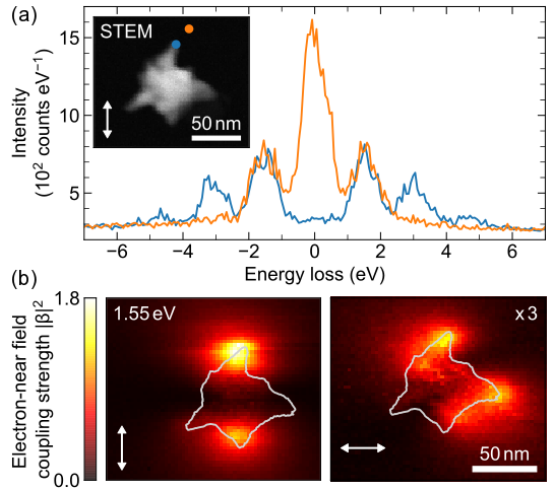

**Fig. 3 |** (a) 200 keV STEM-PINEM spectra of an Au nanostar corresponding to regions of strong (blue curve) and weak (orange curve) electron-near field coupling. The electron beam positions are indicated by the color-matched dots in the inset, showing a HAADF image of the nanostar. The white double arrow represents the approximate in-plane polarization of the driving field. (b) Maps of the electron-near field coupling strength derived from the energy spectra of the transmitted electrons for two orthogonal laser polarizations, as indicated by the white double arrows. The solid lines illustrate the approximate contour of the nanostar as inferred from the HAADF image in (a). Intensities in the right-hand panel have been scaled by a factor of 3.

Next, we map the laser-induced optical field by deriving the electron-near field coupling constant $|\beta|$ from the PINEM spectrum recorded at every electron beam position. To this end, we approximate the electron spectral distribution by a comb of $2N+1$ pseudo-Voigt profiles that are spaced by the photon energy $\hbar\omega_L$, each of which resembling the approximate line shape of the ZLP. The integral of the $n^{\text{th}}$ profile is then determined by the occupation probability $P_n$ of the $n^{\text{th}}$ energy-gain and -loss sideband which again follows from the local coupling constant $|\beta|$ as shown by Eq. (4) (see methods for further details).

The left panel in Fig. 3(b) shows the spatial distribution of the squared electron-near field coupling strength $|\beta|^2$ derived for the same laser polarization as in Fig. 3(a). Clearly, the strongest coupling is observed near the apexes of tips I and II which are both roughly aligned with the laser polarization. Taking the EELS measurements as a reference, this agrees well with the corresponding resonance energies which are closest to the central pump photon energy of 1.55 eV. Indeed, tip III, with its resonance further to the blue, shows almost no response despite a similar symmetry axis as tip II. To verify the correlation with the laser polarization, the right panel of Fig. 3(b) shows a $|\beta|^2$ map for the orthogonal polarization. Clearly, the coupling strength around tips I and II is now strongly reduced (note that the data are scaled by a factor three), and tip IV that is better aligned with the polarization lights up. Yet, the effect is comparably low, consistent with the tip resonance energy being furthest from the central pump photon energy. We note that similar to the polarization, the direction of incidence of the pump field influences the excitation efficiency of a given tip mode. This might contribute to differences in the maximum electron-near field coupling strength observed between tip I and tip II. Three-dimensional tomographic imaging could be used to study this further. Finally, we note that other than EELS, PINEM allows mapping fields when the electrons penetrate through the thicker regions of the nanostar. This is because for the given pump field intensity the electron-near field interaction probability is multiple orders of magnitude larger in PINEM and therefore a larger fraction of electrons undergoes well-defined coherent energy transitions.

**Simulation Results.** To gain quantitative intuition on the EELS, CL, and PINEM response of a nanotip, we perform full-wave retarded BEM calculations for the model geometry introduced in Fig. 1. To simplify the modelling, the effect of a substrate is neglected here[50]. For practical reasons given in Ref. 55, EELS and CL probabilities are calculated using a finite electron beam width of 0.1 nm. For PINEM, the driving field is modelled by a monochromatic plane wave incident from the top and polarized along the symmetry axis of the tip, assuming resonant tip mode excitation at a photon energy of 1.73 eV. We will use our calculations to compare them with experimental data for tip II. The latter has shown a very similar spectral response in EELS and CL, the central pump photon energy in PINEM lies well within the approximate resonance bandwidth of 270 meV, and the tip symmetry axis is roughly oriented in-plane and aligned with the vertical laser polarization chosen in the experiments.

Figure 4 shows 200 keV EELS (a) and 20 keV CL (b) spectra calculated for an electron passing through the center of the spherical core (green curves) and 3 nm away from the tip apex (blue curves), respectively. As expected, for the electron passing near the tip, the tip resonance is clearly observed as a sharp maximum at 1.73 eV with a spectral line width of about 100 meV (FWHM). In the core spectra, the core resonance appears as a broader maximum near 2.4 eV with the small peak at 1.73 eV reflecting coupling to the tip resonance, in good agreement with the experiments. The insets show the two-dimensional EELS and CL distributions around 1.73 eV (bandwidth $\pm 25$ meV). As in the experiments, the maps reveal strong modal field confinement near the tip apex. Comparing the measured and calculated EELS and CL probabilities, we consistently find the EELS probability to be an order of magnitude larger than the CL probability (cf. Fig. 2 (a,b) and Fig. 4 (a,b)). This reflects strong non-radiative losses in the Au plasmonic nanostar at optical frequencies.[11] However, in absolute terms we find the measured EELS and CL probability maxima to be yet an order of magnitude lower than in the calculations. A plausible reason for this deviation is the about two- and four-times larger resonance linewidth found for the tips in EELS and CL, respectively. Among others, this can be explained by a higher damping rate in the metal than predicted by the simulations using optical constants for an extended Au thin film.[56] Spectral broadening also occurs in the measurements due to substrate effects,[50,63] electron-beam induced carbon contamination, and nonlocal effects that could be relevant at the extreme length scale of the tip apexes.[64] As discussed above, in EELS another crucial factor is the finite energy spread in the electron beam that widens the acquired spectra. For CL, we note that light is only collected in the upward hemisphere using a parabolic mirror with limited collection solid angle. In fact, depending on the precise tip orientation, a significant fraction of the radiation will be emitted towards the substrate and will not be detected.[50] A final factor is the electron beam spot size on the sample which can locally influence the signal strength due to spatial averaging in the strongly inhomogeneous plasmonic near fields.

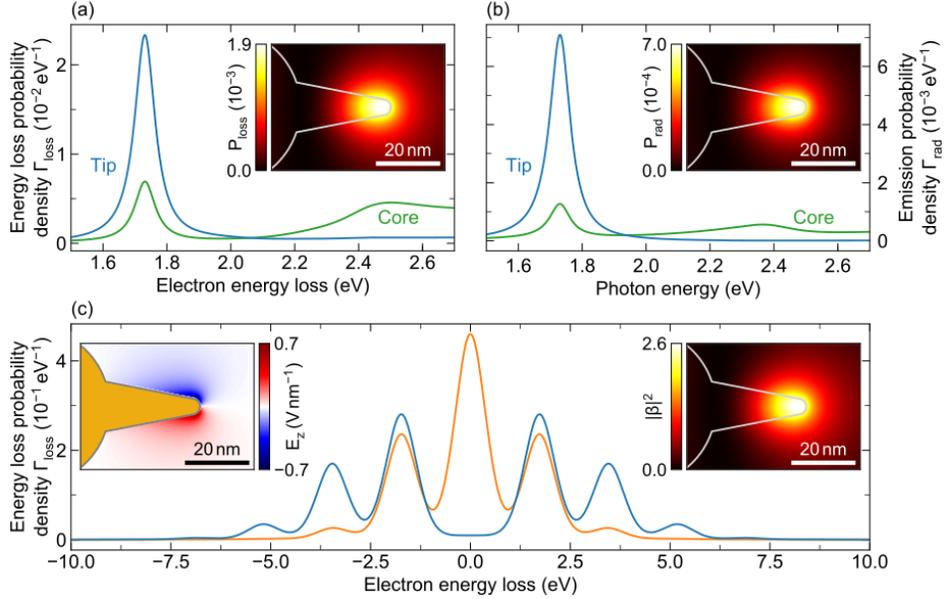

**Fig. 4 |** BEM calculations of (a) 200 keV EELS and (b) 20 keV CL spectra for electrons passing through the center of the spherical core (green) and 3 nm away from the tip apex (blue). The insets show EELS and CL probability distributions obtained for a spectral bandwidth of $\pm 50$ meV around the tip resonance at 1.73 eV. (c) Calculated 200 keV PINEM spectra for electrons passing 3 nm (blue) and 20 nm (orange) away from the tip apex (light plane wave incident along $z$ and polarized along $x$, $\hbar\omega_L$=1.73 eV photon energy, and 20 MW/cm$^2$ light intensity). The left inset shows a time snapshot of the $E_z$ component of the induced near field distribution in the $x$-$z$ plane; the right inset shows a $x$-$y$ map of the calculated electron-photon coupling strength $|\beta|^2$.

In Fig. 4(c) we show 200 keV PINEM spectra calculated for the electron passing by the tip apex at distances of 3 nm (blue curve) and 20 nm (orange curve), respectively. The spectra are represented assuming an equivalent electron beam energy spread as in the experiments (0.9 eV FWHM). For the larger distance, a pronounced ZLP and only the first order sidebands ($\pm \hbar\omega_L$) are observed, while closer to the tip, the ZLP is fully depleted and the first, second, and third order sidebands ($\pm n\hbar\omega_L$ for $n = 1,2,3$) can be seen. In the calculations, a light intensity of 20 MW/cm$^2$ is chosen to best match the electron energy modulation observed in the measurements. This value is about 20 times lower than the experimental pump field intensity of 0.4 GW/cm$^2$ which we assign to slight off-resonant tip excitation, reduced coupling efficiency in the presence of a substrate, imperfect laser alignment relative to the tip symmetry axis (polarization/direction of incidence), deviations in tip morphology as well as larger Ohmic damping than predicted by the calculations (see above). Incidentally, for off-resonant tip excitation at 1.55 eV, a comparable energy modulation is obtained assuming an incident field intensity of 0.25 GW/cm$^2$. We stress that the calculated PINEM spectra match well with the measured ones for both electron-tip separations, indicating similar spatial variations in the near-field intensity. According to Eq. (6), these intensity variations are independent of the driving field intensity and fully determined by the modal field profile, thus enabling a reasonable comparison between numerical and experimental results. For reference, the inset on the left shows the calculated $E_z$ component of the plane wave-induced near field in the $x$-$z$ symmetry plane of the tip; the inset on the right shows the $x$-$y$ distribution of the squared electron-photon interaction strength $|\beta|^2$. The latter shows good correspondence with the experimental data in Fig. 3(b), indicating strong field confinement near the tip apex as in EELS and CL.

**Spatial Dependence.** In EELS and CL, the electrons undergo spontaneous interactions with a number of radiative dipolar tip modes, irrespective of the tip orientation and resonance energy. In contrast, in PINEM the stimulated electron-near field interaction depends on the coupling of these modes to the driving field and thus on its polarization and photon energy. Yet, Eqs. (2,3, and 6) suggest that for an isolated tip mode we can directly compare the spatial distributions retrieved from EELS, CL, and PINEM, and thus quantitatively compare the spatial variations in the spontaneous and stimulated electron-near field interactions at the nanometer length scale.

Figure 5 shows intensity profiles along the symmetry axis of tip II (a-c) and the model nanotip (d). The experimental profiles were obtained by linearly interpolating and averaging the data within the dashed boxes shown in the insets. The EELS and CL data correspond to the fitted loss and emission probability distributions derived from the measurements presented in Fig. 2 at 1.68 eV and 1.72 eV, respectively. For the PINEM profile, data were derived from an additional measurement as in 3(b, left) with increased scanning resolution of (1×1) nm$^2$/pixel. For the model nanotip, we show calculated intensity profiles assuming equivalent conditions as described above. Here, we neglect a near-field contribution from the core resonance due to vanishing spectral overlap with the tip mode.

Following a procedure similar to Ref. 65, we fit the experimental profiles with a model assuming an evanescent exponential decay away from the tip apex with characteristic 1/e decay length $\delta$. The signal along the tip is described by a half-Gaussian distribution peaking at the tip apex, while the finite width of the electron probe is introduced by convolution with a Gaussian resolution function of standard deviation $\sigma$. The fitted curves (dashed lines) and deconvoluted model functions (dotted lines) are plotted in Fig. 4 (a-c). We obtain beam widths of $\sigma_{\text{CL}} = (5.3 \pm 0.2)$ nm for the CL, $\sigma_{\text{EELS}} = (1.6 \pm 0.5)$ nm for the EELS, and $\sigma_{\text{PINEM}} = (2.2 \pm 0.3)$ nm for the PINEM measurements, reflecting the difference in spatial resolution between the employed SEM and STEM instruments. Notably, the beam widths found for EELS and PINEM agree well with the selected STEM probe beam size of 1.5 nm (see methods); for CL a similar value was earlier determined in Ref. 65. The characteristic decay lengths are found to be $\delta_{\text{EELS}} = (8.7 \pm 0.3)$ nm, $\delta_{\text{CL}} = (10.5 \pm 0.2)$ nm, and $\delta_{\text{PINEM}} = (15.2 \pm 0.2)$ nm, showing extreme deep-subwavelength modal field confinement in all cases.

For the calculated profiles we obtain 1/e decay lengths of 8.1 nm for 200 keV EELS and PINEM, and 5.5 nm for 20 keV CL (with respect to the signal amplitude at the tip apex). As expected from Eqs. (2,3, and 6), the calculations yield equivalent decay lengths for the EELS and PINEM profiles, confirming that the spontaneous and stimulated electron-near field interaction processes are purely determined by the spatial field profile of the tip mode, irrespective of the excitation source. The calculated CL profile shows a more rapid decay due a lower electron energy as will be discussed in more detail below. Interestingly, the calculated profiles consistently show that the spontaneous and stimulated electron-near field interaction is strongest for the electrons passing through the tip, about 3 nm nanometers inward from the tip apex. This is in good agreement with the electric field maxima observed in the electron- and laser-induced $E_z$ distributions plotted in Fig. 1(a) and the left-hand inset in Fig. 4(c). The experimental intensity profiles show a very similar functional shape, indicating maximum coupling on the tip.

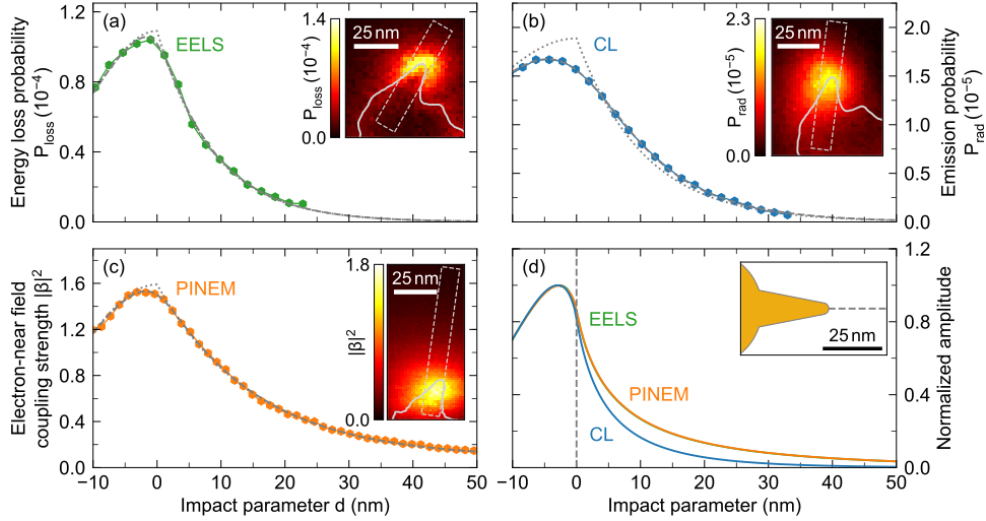

**Fig. 5 |** Line profiles of the probability distributions acquired along tip II using (a) 200 keV EELS (1.68 eV), (b) 20 keV CL (1.72 eV), and 200 keV PINEM ($\hbar\omega_L$ = 1.55 eV photon energy, 0.4 GW/cm² laser intensity). We also plot model fits assuming Gaussian broadening of the data due to the finite width of the electron beam (dashed curves) and deconvoluted model functions (dotted curves). An impact parameter of $d$ = 0 corresponds to the approximate position of the tip apex as derived from the fitting procedure. (d) BEM calculations of the 200 keV EELS (green, hidden behind the PINEM curve), 20 keV CL (blue) and 200 keV PINEM (orange curve) profiles along the tip symmetry axis of the model geometry. The profiles are normalized to their respective amplitudes at the tip apex. As in the experiments, EELS and CL probabilities are calculated over a spectral bandwidth of ±25 meV around the tip resonance at 1.73 eV; the PINEM interaction is calculated for laser polarization along the tip at an excitation energy of $\hbar\omega_L$ = 1.73 eV.

Comparing the experimental and calculated decay lengths, we find excellent agreement for EELS, while for CL and PINEM the experimental values show an upwards deviation of almost 50%. At the scale of a few nanometers, we assume a calibration error on the order of 10-20%. Also, we note that the model nanotip only approximates the actual shape of tip II and thus might show a slightly different near-field extent. For the employed SEM-CL instrument, an accuracy limit of 3 nm was found in previous work,[65] as mostly determined by the large electron probe width. Furthermore, a small uncertainty of 1-2 nm is introduced by the choice of the model to describe the near-field profiles which had to maintain a reasonably small parameter space. The large discrepancy between experimental EELS and PINEM profiles is assigned to sample drift caused by mechanical instabilities or nanoscale heat expansion under laser beam exposure, a prominent effect arising from the relatively long acquisition times that are required to resolve PINEM spectra. This is linked to the low effective beam current which is limited by space-charge related broadening of the electron pulses,[61] also inhibiting drift correction by simultaneous structural imaging. Notably, a line profile through the measured PINEM distribution in Fig. 3(b), taken at larger pixel size and hence shorter acquisition time, yields a decay length of ∼10 nm, close to the values found by EELS and in the calculations. A contribution from the core resonance is expected to be negligibly small as it is spectrally far from the pump field. Overall, we can conclude that in the present case the measurement accuracy of CL and PINEM is practically limited to a few nanometers, however, we stress that this constitutes no fundamental limit and that technical improvements may push the attainable resolution below 1 nm. Importantly, we note that other than EELS, PINEM offers a large SNR even for thicker samples due the high interaction rate while CL requires no electron transparency at all. Moreover, the higher spectral resolution in CL as compared to EELS allows to disentangle the local contribution of modes with spectral overlap much more clearly, in particular in case of tips II and III. We emphasize that similar spectrally-resolved measurements are also possible in PINEM using a tunable pump source.[32]

**Dependence on Electron Energy.** Finally, we analyze the dependence of EELS, CL, and PINEM measurements on the electron energy. As seen above, all three techniques probe the electric field component $E_z$ of the plasmon mode along the electron trajectory. According to Eqs. (3-5), the strength of the measured effect is the result of the field integral along the electron trajectory (i.e., it depends on the Fourier amplitude of the $E_z$ component of the mode at the spatial frequency $q = \omega/v$). Hence, electrons of different energy (velocity) probe different Fourier components. To demonstrate how this affects the electron-near field interaction, we calculate the spatial Fourier composition of the laser-induced $E_z$ distribution plotted in the left inset in Fig. 5(c). Figure 1 shows the Fourier amplitude as a function of along-the-beam wave vector $q$ and distance $d$ away from the tip. Data are expressed in terms of the squared electron-photon interaction strength $|\beta|^2$. The corresponding beam energies (for which the electron velocity is given by $q = \omega/v$) are shown on top.

Several trends can be observed in Fig. 6. First, for a given momentum exchange $\hbar q$, or equivalently, electron energy, the near-field coupling strength rapidly falls off with impact parameter, due to decreasing field intensity. Second, with decreasing distance to the tip, the largest coupling strength is observed at increasing spatial frequencies. This is because the field confinement monotonically increases towards the tip apex. As an example, the inset shows the coupling strength $|\beta|^2$ as a function of wave vector at a distance of 3 nm away from the tip apex. The maximum coupling strength $|\beta|^2$ is observed for the Fourier component with spatial frequency of $q$ = 0.08 nm$^{-1}$, which is best matched by an electron with an energy of 3.5 keV. For increasing distance from the tip, the larger spatial frequency components quickly die out, and the coupling strength peaks for higher electron energy.

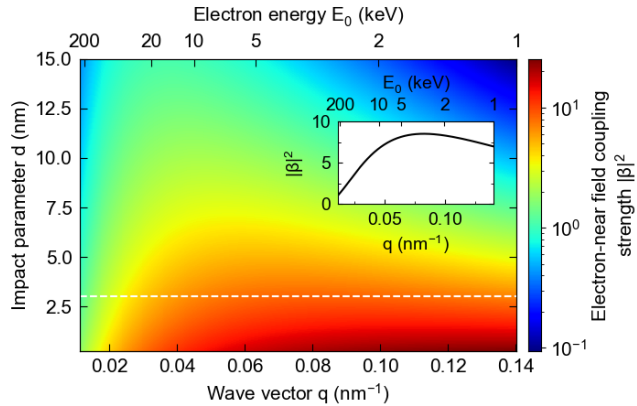

**Fig. 6 |** BEM calculation of the electron-near field coupling strength $|\beta|^2$ as a function of along-the-beam wave vector $q = \omega/v$ and impact parameter $d$ relative to the tip apex, obtained with fixed $\hbar\omega$=1.73 eV tuned to the plasmon energy and varying electron energy (energy $E_0$, top axis). Other excitation parameters are identical as in Fig. 4d and Fig. 5c. The inset shows a profile along the white dashed line for $d$ = 3 nm. For fixed $E_0$, the electron-near field interaction contains a single spatial frequency component, given by the electron-near field momentum exchange $q$.

Overall, we conclude that electrons with kinetic energies of a few keV couple more strongly to the highly confined optical fields at the tip apexes than electrons at typical STEM energies of 80-300 keV. However, these electrons interact more efficiently with more extended areas of the mode at larger distance, or less confined optical fields in dielectric structures. Importantly, Eqs. (2,3 and 6) show that this trend is independent of the spontaneous or stimulated nature of the electron-near field interaction (i.e., it equally applies to the signal strength in EELS, CL, and PINEM). We emphasize that our analysis directly demonstrates how EELS, CL and PINEM are linked to the intrinsic electric field profile of a particular material resonance. Using a more generic approach, the spontaneous electron-near field interaction was recently considered as a scattering problem, i.e. the scattering of the evanescent electron field on a generic dielectric scatterer.[66] Supported by experimental data on a periodic grating structure, it was shown that for sufficiently small electron beam-sample separations the strongest scattering occurs for slow (non-relativistic) electrons as they generate overall larger near-field amplitudes than faster (relativistic) electrons. Here, we explicitly show that depending on the characteristic dimensions of a nanoparticle, the electron-near field coupling peaks for a specific electron energy ranging down to a few keV as can be routinely reached in SEM. Yet, we note that as in case of the SEM-CL instrument employed in this work the low-energy observation of structures as small as the Au nanostars required a trade-off between CL signal strength and spatial resolution due to limitations of the electron optics at low electron energies. Finally, we point out that in general the full reconstruction of the out-of-plane near field component $E_z$ in the spatial domain requires combining multiple EELS, CL, and/or PINEM measurements at different electron energies, such that a range of spatial Fourier components can be probed.

**Conclusion**

In conclusion, we have demonstrated spatially-resolved EELS, CL and PINEM measurements of highly localized optical near fields at the sharp tip apexes of an Au nanostar, enabling a direct comparison of stimulated and spontaneous electron-photon interactions at the nanometer length scale. Our experimental EELS, CL, and PINEM data show good qualitative and quantitative agreement with the results of numerical electromagnetic BEM calculations for a simplified model geometry, taking into account experimental inaccuracies and modelling simplifications. In agreement with theoretical considerations, we find that in the limit of a single isolated tip mode, the electron-near field interaction is independent of the spontaneous or stimulated nature of the process but only varies with the modal electric near-field profile. Specifically, our simulations suggest that the electron-near field interaction is strongest for electrons passing through the tip, a few nanometers inwards from the tip apex, as supported by the experiments. We show that in the proximity of the nanotips, the electron-near field interaction reaches a maximum at electron energies as low as a few keV, depending on the dominant spatial Fourier component in the optical field. Electron microscopy with access to this energy range may thus offer great potential in exploring exciting new phenomena such as quantum-coherent effects in electron-light-matter interactions.

**Methods**

**Sample Preparation.** Au nanostars were prepared by a modification of a previously reported procedure using a seeded growth approach.[42,67] First, spherical Au seeds of approximately 12 nm diameter were produced by a modification of the well-known Turkevich method.[68] The seeds were synthesized by the subsequent addition of dehydrated trisodium citrate ($C_6H_5Na_3O_7 \cdot 2H_2O$, 11 mL, 0.1 M) and gold(III) chloride trihydrate ($HAuCl_4 \cdot 3H_2O$, 833 µL, 0.1 M) to boiling Milli-Q water (500 mL) at intervals of 10 min and under vigorous stirring. After 30 min boiling, the solution was brought to room temperature and the particles were added drop-by-drop under stirring to an aqueous polyvinylpyrrolidone (PVP) solution (500 mL, 0.27 mM). Finally, the Au nanoparticles were centrifuged (9000 rpm, 35 min) and dispersed in ethanol absolute (EtOH, 50 mL) to achieve a final Au concentration of $16.2 \times 10^{-4}$ M. Next, Au nanostars were grown by the fast addition of the PVP-coated Au seeds in EtOH (350 µL) to a PVP solution in N,N-dimethylformamide (DMF, 7 g, 35 mL) containing $HAuCl_4$ (75 µl, 0.12 M aqueous solution) freshly prepared. Within 15 min, the color of the solution turns blue, indicating the formation of Au nanostars. The solution was left under stirring overnight to assure the reduction of all reactants. DMF and excess PVP was removed by several centrifugation steps; a first step at 7500 rpm for 40 min followed by four more iterations at 7000 rpm for 10 min each. For each step, the particles were resuspended in EtOH (35 mL). Eventually, Au nanostars (5 µL, 0.8 mM) were deposited on a TEM silicon nitride support membrane via spin coating (1$^{st}$ ramp: 500 rpm, 10 s; 2$^{nd}$ ramp: 3000 rpm, 30 s with each an acceleration rate of 500 rpm/s) achieving a particle density of approximately 1.2 particles per µm$^2$. In order to minimize contamination issues arising from residue chemicals during exposure to the electron beam, the sample was treated by $O_2$ plasma cleaning for 30 s. PVP (MW = 25 000) was purchased from Roth. $HAuCl_4 \cdot 3H_2O$ (99.9 %), $C_6H_5Na_3O_7 \cdot 2H_2O$ (≥99.5 %), and EtOH (≥99.9%) were obtained from Sigma-Aldrich. DMF (≥99%) was obtained from Fluka. Silicon nitride support membranes were purchased from Ted Pella, Inc. All reactants were used without further purification. Milli-Q water (18 MΩ cm$^{-1}$) was used in all aqueous solutions, and all glassware was cleaned with aqua regia prior to usage.

**EELS and PINEM measurements** were performed in STEM mode of a JEOL JEM-2100F TEM instrument based on a custom modified Schottky field emission source, with a selected electron-probe beam diameter of 1.5 nm. The spectral scans were recorded with a CEOS energy filtering and imaging device (CEFID) equipped with a TVIPS TemCam-XF416ES scintillator-coupled CMOS camera and synchronized by a TVIPS universal scan generator. For EELS, the full dispersion of the spectrometer was set to 16 eV over 4096 pixels at 4-fold binning, corresponding to a bin size of 15.6 meV. For PINEM, a dispersion of 68 eV over 4096 pixels was used, resulting in a bin size of 16.6 meV. EELS spectra were acquired using a continuous electron beam with an initial energy spread of 0.5 eV (ZLP FHWM). For PINEM, the instrument was operated in ultrafast laser-triggered photoemission configuration enabling synchronous sample exposure by sub-picosecond electron probe and picosecond optical pump pulses. An amplified Ti:Saphire laser system (Coherent RegA) provided femtosecond pulses at a central photon energy of 1.55 eV ($\lambda$ = 800 nm) and a spectral bandwidth of 65 meV (35-nm bandwidth), at 600 kHz repetition rate. The optical pump pulses were dispersively stretched to 3.4-ps pulse duration in a 19 cm bar of dense flint glass (SF6). The sample was excited under near-normal incidence (parallel to the electron beam) and controllable polarization state with the light focused to a spot diameter of ~15 μm and an average power of 4 mW (corresponding to a peak intensity of about 0.4 GW/cm$^2$). Synchronous sub-picosecond electron-probe pulses were generated by photoemission from the Schottky field emitter using the second harmonic of the fundamental laser beam (for further details see Ref. 61). The energy spread of the electron pulses was about 0.9 eV. PINEM and EELS spectra were acquired at an integration time of 500 ms and 120 ms, respectively. The EELS spectra were deconvoluted by a Richardson-Lucy algorithm[58] implemented in the Hyperspy Python library[69] using 20 iterations. Absolute EELS probabilities were obtained by normalizing the spectra to the integrated number of counts measured upon electron beam transmission through the silicon nitride support membrane.

**CL measurements** were performed in a Thermo Fisher Scientific/FEI Quanta FEG 650 SEM equipped with a Schottky field emission electron source operated a beam current of about 570 pA. CL emission was collected by a half-parabolic mirror covering a solid angle of 1.46 $\pi$ steradians above the sample plane, and directed into a DELMIC SPARC optical detection system for spectrally-resolved CL analysis.[70] The acquisition time for each spectrum was 350 ms and the resolution of the spectrometer was in the order of 10 meV (as determined from the sharp emission lines of an argon calibration lamp). Secondary electron images were taken simultaneously to the CL acquisition and software-controlled drift correction was applied at time intervals of 1 s in order to compensate for the effect of mechanical instabilities or electrostatic charging. Background luminescence from the silicon nitride support membrane was measured separately and subtracted from the raw CL data. The system response was calibrated and absolute CL probabilities were obtained using the transition radiation (TR) spectrum measured upon 20 keV electron beam impact on the flat surface of a single-crystalline Al sample normalized to the analytically calculated TR spectrum using the expression given in Ref. 2 with optical material constants derived from spectroscopic ellipsometry data.

**Analysis of the EELS and CL spectra.** In a first step, the EELS and CL spectra are averaged over segments of 10×10 pixels, and a global fit is performed to determine the central energy $E_0$ and linewidth $\gamma$ (FWHM) of the tip (and core) resonances. In this procedure the resonance amplitudes and background parameters constitute local fit parameters. Averaging the data improves convergence of the minimization process while at the same time ensures sufficient signal-to-noise ratio. In a second step, the amplitude of each resonance is fitted to the spectrum taken at every electron beam position given the fixed resonance energies and linewidths obtained previously.

**Analysis of the PINEM Spectra.** The derivation of the electron-near field coupling constant from the PINEM spectra is done following a similar procedure as described in the supplementary information to Ref. 39. The initial electron energy distribution (i.e., prior to the near field interaction) is modeled by a comb of pseudo-Voigt profiles with a Lorentzian-like contribution of 25% and a FWHM of 0.9 eV. Furthermore, we assume a Gaussian distribution of the coupling constant $\beta$ with a standard deviation of $\Delta\beta/\beta = 0.2$ in order to account for residual spatial and temporal averaging in the strongly inhomogeneous optical near field, motivated by the finite width of the electron beam, as well as the temporal profile of the optical pump pulses.

*Conflict of interest.* The authors declare the following competing financial interests: A.P. is co-founder and co-owner of Delmic BV, a company that produces the cathodoluminescence system that was used in this work.

*Acknowledgements.* This project has received funding from the European Research Council (ERC) under the European Union's Horizon 2020 research and innovation programme (grant agreement No. 695343). Work at AMOLF is partly financed by the Dutch Research Council (NWO). Work at the University of Göttingen was funded by the Deutsche Forschungsgemeinschaft (DFG, German Research Foundation) (217133147/SFB 1073 project A05 and 255652344/SPP 1840 project 'Kohärente Wechselwirkungen starker optischer Nahfelder mit freien Elektronen'), and the Gottfried Wilhelm Leibniz program. Work at URV was financed by Spanish Ministerio de Economia y Competitividad (MINECO) (CTQ2017-88648R and RYC-2015-19107), the Generalitat de Cataluña (2017SGR883), the Universitat Rovira i Virgili (2018PFR-URV-B2-02), and the Banco Santander (2017EXIT-08). JGA received funding from the ERC (Advanced Grant No. 789104-eNANO), Spanish MINECO (MAT2017-88492-R and SEV2015-0522), Catalan CERCA Program, and Fundació Privada Cellex.

**References**

(1)  Koenderink, A. F.; Polman, A. Nanophotonics: Shrinking Light-Based Technology. *Science.* **2015**, *348* (6234), 516–521.

(2)  García De Abajo, F. J. Optical Excitations in Electron Microscopy. *Rev. Mod. Phys.* **2010**, *82* (1), 209–275.

(3)  Losquin, A.; Lummen, T. T. A. Electron Microscopy Methods for Space-, Energy-, and Time-Resolved Plasmonics. *Front. Phys.* **2017**, *12* (1), 1–27.

(4)  Polman, A.; Kociak, M.; García de Abajo, F. J. Electron-Beam Spectroscopy for Nanophotonics. *Nat. Mater.* **2019**, *18* (11), 1158–1171. https://doi.org/10.1038/s41563-019-0409-1.

(5)  Brenny, B. J. M.; Polman, A.; García De Abajo, F. J. Femtosecond Plasmon and Photon Wave Packets Excited by a High-Energy Electron on a Metal or Dielectric Surface. *Phys. Rev. B* **2016**, *94* (15), 1–11.

(6)  Wu, Y.; Li, G.; Camden, J. P. Probing Nanoparticle Plasmons with Electron Energy Loss Spectroscopy. *Chem. Rev.* **2018**, *118* (6), 2994–3031.

(7)  Colliex, C.; Kociak, M.; Stéphan, O. Electron Energy Loss Spectroscopy Imaging of Surface Plasmons at the Nanometer Scale. *Ultramicroscopy* **2016**, *162*, A1–A24.

(8)  Coenen, T.; Brenny, B. J. M.; Vesseur, E. J.; Polman, A. Cathodoluminescence Microscopy: Optical Imaging and Spectroscopy with Deep-Subwavelength Resolution. *MRS Bull.* **2015**, *40* (4), 359–365.

(9)  García De Abajo, F. J.; Kociak, M. Probing the Photonic Local Density of States with Electron Energy Loss Spectroscopy. *Phys. Rev. Lett.* **2008**, *100* (10), 1–4.

(10) Losquin, A.; Kociak, M. Link between Cathodoluminescence and Electron Energy Loss Spectroscopy and the Radiative and Full Electromagnetic Local Density of States. *ACS Photonics* **2015**, *2* (11), 1619–1627.

(11) Losquin, A.; Zagonel, L. F.; Myroshnychenko, V.; Rodríguez-González, B.; Tencé, M.; Scarabelli, L.; Förstner, J.; Liz-Marzán, L. M.; García De Abajo, F. J.; Stéphan, O.; Kociak, M. Unveiling Nanometer Scale Extinction and Scattering Phenomena through Combined Electron Energy Loss Spectroscopy and Cathodoluminescence Measurements. *Nano Lett.* **2015**, *15* (2), 1229–1237.

(12) Nelayah, J.; Kociak, M.; Stéphan, O.; De Abajo, F. J. G.; Tencé, M.; Henrard, L.; Taverna, D.; Pastoriza-Santos, I.; Liz-Marzán, L. M.; Colliex, C. Mapping Surface Plasmons on a Single Metallic Nanoparticle. *Nat. Phys.* **2007**, *3* (5), 348–353.

(13) Schaffer, B.; Hohenester, U.; Trügler, A.; Hofer, F. High-Resolution Surface Plasmon Imaging of Gold Nanoparticles by Energy-Filtered Transmission Electron Microscopy. *Phys. Rev. B - Condens. Matter Mater. Phys.* **2009**, *79* (4), 1–4.

(14) Coenen, T.; Van De Groep, J.; Polman, A. Resonant Modes of Single Silicon Nanocavities Excited by Electron Irradiation. *ACS Nano* **2013**, *7* (2), 1689–1698.

(15) Coenen, T.; Bernal Arango, F.; Femius Koenderink, A.; Polman, A. Directional Emission from a Single Plasmonic Scatterer. *Nat. Commun.* **2014**, *5*, 1–8.

(16) Coenen, T.; Schoen, D. T.; Brenny, B. J. M.; Polman, A.; Brongersma, M. L. Combined Electron Energy-Loss and Cathodoluminescence Spectroscopy on Individual and Composite Plasmonic Nanostructures. *Phys. Rev. B* **2016**, *93* (19).

(17) Mignuzzi, S.; Mota, M.; Coenen, T.; Li, Y.; Mihai, A. P.; Petrov, P. K.; Oulton, R. F. M.; Maier, S. A.; Sapienza, R. Energy-Momentum Cathodoluminescence Spectroscopy of Dielectric Nanostructures. *ACS Photonics* **2018**, *5* (4), 1381–1387.

(18) Raza, S.; Stenger, N.; Pors, A.; Holmgaard, T.; Kadkhodazadeh, S.; Wagner, J. B.; Pedersen, K.; Wubs, M.; Bozhevolnyi, S. I.; Mortensen, N. A. Extremely Confined Gap Surface-Plasmon Modes Excited by Electrons. *Nat. Commun.* **2014**, *5* (May), 1–7.

(19) Raza, S.; Esfandyarpour, M.; Koh, A. L.; Mortensen, N. A.; Brongersma, M. L.; Bozhevolnyi, S. I. Electron Energy-Loss Spectroscopy of Branched Gap Plasmon Resonators. *Nat. Commun.* **2016**, *7* (1), 1–10.

(20) Vesseur, E. J. R.; Coenen, T.; Caglayan, H.; Engheta, N.; Polman, A. Experimental Verification of N=0 Structures for Visible Light. *Phys. Rev. Lett.* **2013**, *110* (1), 1–5.

(21) Sapienza, R.; Coenen, T.; Renger, J.; Kuttge, M.; Van Hulst, N. F.; Polman, A. Deep-Subwavelength Imaging of the Modal Dispersion of Light. *Nat. Mater.* **2012**, *11* (9), 781–787.

(22) Peng, S.; Schilder, N. J.; Ni, X.; Van De Groep, J.; Brongersma, M. L.; Alù, A.; Khanikaev, A. B.; Atwater, H. A.; Polman, A. Probing the Band Structure of Topological Silicon Photonic Lattices in the Visible Spectrum. *Phys. Rev. Lett.* **2019**, *122* (11), 117401.

(23) Barwick, B.; Flannigan, D. J.; Zewail, A. H. Photon-Induced near-Field Electron Microscopy. *Nature* **2009**, *462* (7275), 902–906.

(24) García de Abajo, F. J.; Kociak, M. Electron Energy-Gain Spectroscopy. *New J. Phys.* **2008**, *10* (7), 073035.

(25) Park, S. T.; Lin, M.; Zewail, A. H. Photon-Induced near-Field Electron Microscopy (PINEM): Theoretical and Experimental. *New J. Phys.* **2010**, *12*.

(26) Garcia De Abajo, F. J.; Asenjo-Garcia, A.; Kociak, M. Multiphoton Absorption and Emission by Interaction of Swift Electrons with Evanescent Light Fields. *Nano Lett.* **2010**, *10* (5), 1859–1863.

(27) Di Giulio, V.; Kociak, M.; de Abajo, F. J. G. Probing Quantum Optical Excitations with Fast Electrons. *Optica* **2019**, *6* (12), 1524.

(28) Yurtsever, A.; Zewail, A. H. Direct Visualization of Near-Fields in Nanoplasmonics and Nanophotonics. *Nano Lett.* **2012**, *12* (6), 3334–3338.

(29) Feist, A.; Echternkamp, K. E.; Schauss, J.; Yalunin, S. V.; Schäfer, S.; Ropers, C. Quantum Coherent Optical Phase Modulation in an Ultrafast Transmission Electron Microscope. *Nature* **2015**, *521* (7551), 200–203.

(30) Piazza, L.; Lummen, T. T. A.; Quiñonez, E.; Murooka, Y.; Reed, B. W.; Barwick, B.; Carbone, F. Simultaneous Observation of the Quantization and the Interference Pattern of a Plasmonic Near-Field. *Nat. Commun.* **2015**, *6*.

(31) Morimoto, Y.; Baum, P. Diffraction and Microscopy with Attosecond Electron Pulse Trains. *Nat. Phys.* **2018**, *14* (3), 252–256.

(32) Pomarico, E.; Madan, I.; Berruto, G.; Vanacore, G. M.; Wang, K.; Kaminer, I.; García De Abajo, F. J.; Carbone, F. MeV Resolution in Laser-Assisted Energy-Filtered Transmission Electron Microscopy. *ACS Photonics* **2018**, *5* (3), 759–764.

(33) Wang, K.; Dahan, R.; Shentcis, M.; Kauffmann, Y.; Ben Hayun, A.; Reinhardt, O.; Tsesses, S.; Kaminer, I. Coherent Interaction between Free Electrons and a Photonic Cavity. *Nature* **2020**, *582* (7810), 50–54.

(34) Nehemia, S.; Dahan, R.; Shentcis, M.; Reinhardt, O.; Adiv, Y.; Wang, K.; Beer, O.; Kurman, Y.; Shi, X.; Lynch, M. H.; Kaminer, I. Observation of the Stimulated Quantum Cherenkov Effect. **2019**, ArXiv: 1909.00757.

(35) Kfir, O.; Lourenço-Martins, H.; Storeck, G.; Sivis, M.; Harvey, T. R.; Kippenberg, T. J.; Feist, A.; Ropers, C. Controlling Free Electrons with Optical Whispering-Gallery Modes. *Nature* **2020**, *582* (7810), 46–49.

(36) Priebe, K. E.; Rathje, C.; Yalunin, S. V.; Hohage, T.; Feist, A.; Schäfer, S.; Ropers, C. Attosecond Electron Pulse Trains and Quantum State Reconstruction in Ultrafast Transmission Electron Microscopy. *Nat. Photonics* **2017**, *11* (12), 793–797.

(37) Vanacore, G. M.; Madan, I.; Berruto, G.; Wang, K.; Pomarico, E.; Lamb, R. J.; McGrouther, D.; Kaminer, I.; Barwick, B.; García De Abajo, F. J.; Carbone, F. Attosecond Coherent Control of Free-Electron Wave Functions Using Semi-Infinite Light Fields. *Nat. Commun.* **2018**, *9* (1).

(38) Vanacore, G. M.; Berruto, G.; Madan, I.; Pomarico, E.; Biagioni, P.; Lamb, R. J.; McGrouther, D.; Reinhardt, O.; Kaminer, I.; Barwick, B.; Larocque, H.; Grillo, V.; Karimi, E.; García de Abajo, F. J.; Carbone, F. Ultrafast Generation and Control of an Electron Vortex Beam via Chiral Plasmonic near Fields. *Nat. Mater.* **2019**, *18* (6), 573–579.

(39) Harvey, T. R.; Henke, J. W.; Kfir, O.; Lourenço-Martins, H.; Feist, A.; García De Abajo, F. J.; Ropers, C. Probing Chirality with Inelastic Electron-Light Scattering. *Nano Lett.* **2020**, *20* (6), 4377–4383.

(40) Asenjo-Garcia, A.; García De Abajo, F. J. Plasmon Electron Energy-Gain Spectroscopy. *New J. Phys.* **2013**, *15*.

(41) Das, P.; Blazit, J. D.; Tencé, M.; Zagonel, L. F.; Auad, Y.; Lee, Y. H.; Ling, X. Y.; Losquin, A.; Colliex, C.; Stéphan, O.; García de Abajo, F. J.; Kociak, M. Stimulated Electron Energy Loss and Gain in an Electron Microscope without a Pulsed Electron Gun. *Ultramicroscopy* **2019**, *203* (December 2018), 44–51.

(42) Pazos-Perez, N.; Guerrini, L.; Alvarez-Puebla, R. A. Plasmon Tunability of Gold Nanostars at the Tip Apexes. *ACS Omega* **2018**, *3* (12), 17173–17179.

(43) Hao, E.; Bailey, R. C.; Schatz, G. C.; Hupp, J. T.; Li, S. Synthesis and Optical Properties of "Branched" Gold Nanocrystals. *Nano Lett.* **2004**, *4* (2), 327–330.

(44) Nehl, C. L.; Liao, H.; Hafner, J. H. Optical Properties of Star-Shaped Gold Nanoparticles. *Nano Lett.* **2006**, *6* (4), 683–688.

(45) Hao, F.; Nehl, C. L.; Hafner, J. H.; Nordlander, P. Plasmon Resonances of a Gold Nanostar. *Nano Lett.* **2007**, *7* (3), 729–732.

(46) Senthil Kumar, P.; Pastoriza-Santos, I.; Rodríguez-González, B.; Javier García De Abajo, F.; Liz-Marzán, L. M. High-Yield Synthesis and Optical Response of Gold Nanostars. *Nanotechnology* **2008**, *19* (1).

(47) Hrelescu, C.; Sau, T. K.; Rogach, A. L.; Jäckel, F.; Laurent, G.; Douillard, L.; Charra, F. Selective Excitation of Individual Plasmonic Hotspots at the Tips of Single Gold Nanostars. *Nano Lett.* **2011**, *11* (2), 402–407.

(48) Mazzucco, S.; Stéphan, O.; Colliex, C.; Pastoriza-Santos, I.; Liz-Marzan, L. M.; De Abajo, J. G.; Kociak, M. Spatially Resolved Measurements of Plasmonic Eigenstates in Complex-Shaped, Asymmetric Nanoparticles: Gold Nanostars. *EPJ Appl. Phys.* **2011**, *54* (3), 1–9.

(49) Shao, L.; Susha, A. S.; Cheung, L. S.; Sau, T. K.; Rogach, A. L.; Wang, J. Plasmonic Properties of Single Multispiked Gold Nanostars: Correlating Modeling with Experiments. *Langmuir* **2012**, *28* (24), 8979–8984.

(50) Das, P.; Kedia, A.; Kumar, P. S.; Large, N.; Chini, T. K. Local Electron Beam Excitation and Substrate Effect on the Plasmonic Response of Single Gold Nanostars. *Nanotechnology* **2013**, *24* (40), 1–8.

(51) Maity, A.; Maiti, A.; Das, P.; Senapati, D.; Chini, T. K. Effect of Intertip Coupling on the Plasmonic Behavior of Individual Multitipped Gold Nanoflower. *ACS Photonics* **2014**, *1* (12), 1290–1297.

(52) Sivis, M.; Pazos-Perez, N.; Yu, R.; Alvarez-Puebla, R.; García de Abajo, F. J.; Ropers, C. Continuous-Wave Multiphoton Photoemission from Plasmonic Nanostars. *Commun. Phys.* **2018**, *1* (1), 3–8.

(53) García de Abajo, F. J.; Howie, A. Retarded Field Calculation of Electron Energy Loss in Inhomogeneous Dielectrics. *Phys. Rev. B - Condens. Matter Mater. Phys.* **2002**, *65* (11), 1154181–11541817.

(54) Hohenester, U.; Trügler, A. MNPBEM - A Matlab Toolbox for the Simulation of Plasmonic Nanoparticles. *Comput. Phys. Commun.* **2012**, *183* (2), 370–381.

(55) Hohenester, U. Simulating Electron Energy Loss Spectroscopy with the MNPBEM Toolbox. *Comput. Phys. Commun.* **2014**, *185* (3), 1177–1187.

(56) P. B. Johnson and R. W. Christy. Optical Constant of the Nobel Metals. *Phys. L Re View B* **1972**, *6* (12), 4370–4379.

(57) Bosman, M.; Ye, E.; Tan, S. F.; Nijhuis, C. A.; Yang, J. K. W.; Marty, R.; Mlayah, A.; Arbouet, A.; Girard, C.; Han, M. Y. Surface Plasmon Damping Quantified with an Electron Nanoprobe. *Sci. Rep.* **2013**, *3*, 1–7.

(58) Gloter, A.; Douiri, A.; Tencé, M.; Colliex, C. Improving Energy Resolution of EELS Spectra: An Alternative to the Monochromator Solution. *Ultramicroscopy* **2003**, *96* (3–4), 385–400.

(59) Kawasaki, N.; Meuret, S.; Weil, R.; Lourenço-Martins, H.; Stéphan, O.; Kociak, M. Extinction and Scattering Properties of High-Order Surface Plasmon Modes in Silver Nanoparticles Probed by Combined Spatially Resolved Electron Energy Loss Spectroscopy and Cathodoluminescence. *ACS Photonics* **2016**, *3* (9), 1654–1661.

(60) Hettler, S.; Dries, M.; Hermann, P.; Obermair, M.; Gerthsen, D.; Malac, M. Carbon Contamination in Scanning Transmission Electron Microscopy and Its Impact on Phase-Plate Applications. *Micron* **2017**, *96*, 38–47.

(61) Feist, A.; Bach, N.; Rubiano da Silva, N.; Danz, T.; Möller, M.; Priebe, K. E.; Domröse, T.; Gatzmann, J. G.; Rost, S.; Schauss, J.; Strauch, S.; Bormann, R.; Sivis, M.; Schäfer, S.; Ropers, C. Ultrafast Transmission Electron Microscopy Using a Laser-Driven Field Emitter: Femtosecond Resolution with a High Coherence Electron Beam. *Ultramicroscopy* **2017**, *176* (November 2016), 63–73.

(62) Bach, N.; Domröse, T.; Feist, A.; Rittmann, T.; Strauch, S.; Ropers, C.; Schäfer, S. Coulomb Interactions in High-Coherence Femtosecond Electron Pulses from Tip Emitters. *Struct. Dyn.* **2019**, *6* (1).

(63) Vernon, K. C.; Funston, A. M.; Novo, C.; Gómez, D. E.; Mulvaney, P.; Davis, T. J. Influence of Particle-Substrate Interaction on Localized Plasmon Resonances. *Nano Lett.* **2010**, *10* (6), 2080–2086.

(64) Myroshnychenko, V.; Rodríguez-Fernández, J.; Pastoriza-Santos, I.; Funston, A. M.; Novo, C.; Mulvaney, P.; Liz-Marzán, L. M.; García De Abajo, F. J. Modelling the Optical Response of Gold Nanoparticles. *Chem. Soc. Rev.* **2008**, *37* (9), 1792–1805.

(65) Schefold, J.; Meuret, S.; Schilder, N.; Coenen, T.; Agrawal, H.; Garnett, E. C.; Polman, A. Spatial Resolution of Coherent Cathodoluminescence Super-Resolution Microscopy. *ACS Photonics* **2019**, *6* (4), 1067–1072.

(66) Yang, Y.; Massuda, A.; Roques-Carmes, C.; Kooi, S. E.; Christensen, T.; Johnson, S. G.; Joannopoulos, J. D.; Miller, O. D.; Kaminer, I.; Soljačić, M. Maximal Spontaneous Photon Emission and Energy Loss from Free Electrons. *Nat. Phys.* **2018**, *14* (9), 894–899.

(67) Pastoriza-santos, I.; Rodrı, L.; Mazzucco, S.; Ste, O.; Kociak, M.; Liz-marza, L. M.; Garcı, F. J.; Rodríguez-Lorenzo, L.; Álvarez-Puebla, R. A.; Stéphan, O.; Liz-Marzán, L. M.; García de Abajo, F. J. Zeptomol Detection Through Controlled Ultrasensitive Surface-Enhanced Raman Scattering. *J. Am. Chem. Soc.* **2009**, *131* (13), 4616–4618.

(68) Turkevich, J. Colloidal Gold. Part II. *Gold Bull.* **1985**, *18* (4), 125–131.

(69) de la Peña, F.; Ostasevicius, T.; Fauske, V. T.; Burdet, P.; Jokubauskas, P.; Nord, M.; Prestat, E.; Sarahan, M.; MacArthur, K. E.; Johnstone, D. N.; Taillon, J.; Caron, J.; Furnival, T.; Eljarrat, A.; Mazz, S. Hyperspy 1.3. 2017.

(70) Coenen, T.; den Hoedt, S. V.; Polman, A. A New Cathodoluminescence System for Nanoscale Optics, Materials Science, and Geology. *Micros. Today* **2016**, *24* (3), 12–19.